\documentclass[]{aastex631}

\usepackage{xcolor}

\begin{document}

\title{Memory in the Burst Occurrence of Repeating FRBs}

\author[0000-0003-0466-2223]{Ping Wang}
\email{pwang@ihep.ac.cn}

\author{Li-Ming Song}
\author{Shao-Lin Xiong}
\author{Xiao-Yun Zhao}
\author{Jin Wang}
\affiliation{Institute of High Energy Physics, Chinese Academy of Science, 100049 Beijing, China}

\author{Shu-Min Zhao}
\affiliation{Institute of High Energy Physics, Chinese Academy of Science, 100049 Beijing, China}
\affiliation{Tangshan Research Institute, Southwest Jiaotong University,  063002 Tangshan, China}

\author{Shuo Xiao}
\affiliation{School of Physics and Electronic Science, Guizhou Normal University, 550001 Guiyang, China}

\author{Ce Cai}
\affiliation{College of Physics and Hebei Key Laboratory of Photophysics Research and Application, Hebei Normal University, Shijiazhuang, 050024 Hebei, China}

\author{Sheng-Lun Xie}
\affiliation{Institute of High Energy Physics, Chinese Academy of Science, 100049 Beijing, China}
\affiliation{Institute of Astrophysics, Central China Normal University, 430000, Wuhan, China}

\author{Wang-Chen Xue}
\author{Chen-Wei Wang}
\author{Yue Wang}
\affiliation{Institute of High Energy Physics, Chinese Academy of Science, 100049 Beijing, China}

\author{Wen-Long Zhang}
\affiliation{Institute of High Energy Physics, Chinese Academy of Science, 100049 Beijing, China}
\affiliation{School of Physics and Physical Engineering, Qufu Normal University, 273165 Qufu, China}

\correspondingauthor{Ping Wang}



\begin{abstract}

Understanding the nature of repeating fast radio bursts (FRBs) is crucial to probe their underlying physics. In this work, we analyze the waiting time statistics between bursts of three repeating FRBs from four datasets. 
We find a universally pronounced dependency of the waiting times on the previous time interval (denoted as $\lambda_0$). We observe a temporal clustering where short waiting times tend to be followed by short ones, and long by long comparative to their mean value. This memory dependency is manifested in the conditional mean waiting time as well as in the conditional mean residual time to the next burst, both of which increase in direct proportion to $\lambda_0$. Consequently, the likelihood of experiencing a subsequent FRB burst within a given time window after the preceding burst is generally influenced by the burst history. We reveal that for the first time, these memory effects are present in the scale-invariant preconditioned waiting time distribution. We show that the memory effect provides a unified description of waiting times which may account for both the repeating FRBs and the apparent non-repeating FRBs (i.e. only observed one time). These results shed new light on the mechanism of FRBs.

\end{abstract}

\keywords{Radio bursts (1339) ---  Radio transient sources (2008)}


\section{Introduction} \label{sec:intro}

Fast Radio Bursts (FRBs) represent one of the most powerful astronomical phenomena discovered in recent years \citep{10.1146/annurev-astro-091918-104501, bailesDiscoveryScientificPotential2022}. Characterized by their brief yet energetic emissions, lasting merely a few milliseconds, they release energy comparable to what the sun emits over several days. The study of FRBs holds the promise of unveiling critical insights into the cosmic medium, the distribution of matter among galaxies and galaxy clusters \citep{10.3847/0004-637x/824/2/105, 10.1007/s00159-019-0116-6}. Moreover, observing the delay and frequency drift in FRBs offers valuable insights into the nature and origin of these mysterious cosmic phenomena \citep{10.3847/2041-8213/ab13ae, 10.1093/mnras/stz700}, which can help constrain theoretical models and shed light on the mechanisms responsible for their generation. These investigations bear significant physical implications and contribute to our understanding of fundamental astrophysical processes. One of the most significant discoveries in the study of FRBs is the identification of a repeating FRB known as FRB 20121102A \citep{10.1038/nature17168}. This repeating burst has provided valuable insights into the physical nature of FRBs and has supported models that propose an extragalactic neutron star as the origin of these bursts \citep{Hessels_2017, Josephy_2019, 10.1007/s11467-020-1039-4, 10.1093/mnras/stac2596}.

However, the origin and propagation mechanisms surrounding FRBs remain mysterious, inspiring an abundance of theories. Among the proposed models, the collision or merger of compact objects like neutron stars or black holes, and the activity of highly magnetized neutron stars, known as magnetars, have attracted substantial attention. The discovery of X-ray emissions accompanying an FRB from a magnetar in the Milky Way, for instance, hints at the plausible role of magnetars in FRB production \citep{ bochenekFastRadioBurst2020a, andersenBrightMilliseconddurationRadio2020, liHXMTIdentificationNonthermal2021}. Furthermore, the young, highly magnetized, extragalactic neutron star model, supported by the observation of FRB 20121102A, has also found support within possible models \citep{10.1038/nature17168}.

Despite the progress made in understanding their physical properties through the study of host galaxies, current models do not fully explain all aspects of FRBs \citep{10.1038/s41586-020-2828-1, zhangPhysicsFastRadio2023, kirstenLinkRepeatingNonrepeating2024}. In particular, the phenomenon of repeating bursts, in which multiple bursts emerge from the same source, remains a substantial enigma. Although some FRBs are observed to repeat, others appear as one-off events without subsequent detections. This contrast raises intriguing questions about the nature of these bursts. The presence of non-repeating FRBs is clear; however, it remains an open question whether these observed non-repetitive bursts might still exhibit repetitive behavior under extended observation periods or with improved detection technologies. As noted in \cite{zhangPhysicsFastRadio2023}, the hypothesis that all FRBs could potentially repeat has not been definitively ruled out and requires further investigation. Supporting this notion, 
\cite{kirstenLinkRepeatingNonrepeating2024} provide empirical evidence suggesting that high-energy burst distributions of certain hyperactive repeating sources resemble those of the non-repeating FRB population. This observation indicates that apparently non-repeating FRB sources may simply be the rarest bursts from repeating sources, which could be observable across different energy scales.
Continued research into the intermittent phenomena and the extended time intervals between these FRBs, as exhibited in existing records, is essential to develop a more complete understanding of FRB activity. The findings of \cite{kirstenLinkRepeatingNonrepeating2024} underscore the complexity of categorizing FRBs definitively as repeaters or non-repeaters based on current observational capabilities and highlight the need for sustained observational efforts across multiple energy thresholds.

Since the identification of the first FRB in 2007 \citep{lorimerBrightMillisecondRadio2007}, the tally of discovered FRBs has increased to more than 700, showing over 60 instances of repeated burst phenomena \citep{Blinkverse}. The ongoing debate on the fundamental difference, if any, between the origins and physical mechanisms of repeating and non-repeating bursts continues to stimulate investigation. Recent efforts by FAST, involving continuous observations of several repeating FRBs with low threshold and high time resolution, have collected a wealth of data on repeating FRBs. This rich repository of information provides an opportunity to uncover more clues about the origin mechanisms of FRBs and the key temporal characteristics of repeating bursts through rigorous statistical analysis.

The temporal analysis of repeating FRBs suggests an evolving nature of their sources over time. If the emission of FRBs followed a Poisson process, the occurrence of bursts would be random, with a consistent number within any given time frame. However, recent research has considered the likelihood of periodicity in the timing of FRB events \citep{frbcollaborationSubsecondPeriodicityFast2022a, weiPeriodicOriginFast2022}, and evidence points to a non-Poisson distribution in some repeating FRBs \citep{wang_sgr-like_2017, Oppermann2018, wangRepeatingFastRadio2023, duScalingUniversalityTemporal2024a}. A newly identified characteristic, demonstrated by the scaling properties in burst timing across diverse sources, persists despite the variable environments of these sources \citep{duScalingUniversalityTemporal2024a}. The uniformity observed in time interval distributions from different FRB sources, even when influenced by varying fluences, emphasizes the significance of interevent time statistics in understanding the dynamics of bursts. Contrasting with the distribution of burst energy, these findings could underline the potential to uncover key dynamics of the burst mechanism through statistical examination of the event time distributions \citep{kumarIntereventTimeDistributions2020}.

The specific distribution function, scaled by the average interval time of the bursts $\bar{\lambda}$, is expressed as
\begin{equation}
P_I(\lambda) \sim \frac{1}{\bar{\lambda}_I} f(\lambda/\bar{\lambda}_I). \label{scale_law}
\end{equation}
The mean waiting time, denoted as $\bar{\lambda}$ is dependent on magnitude $I$ of fluence and defined during continuous observing periods. This distribution closely approximates the function commonly known as the generalized Pareto distribution \citep{hosking_parameter_1987},
\begin{equation}
f(\tau) \sim (1 + \frac{\tau}{\beta})^{-\gamma}, \label{g_distribution}
\end{equation}
where $\beta$ and $\gamma$ are fitting parameters. Specifically, for $\gamma$ as the shape parameter, which determines how the tail of the distribution decays. The Pareto distribution, known for its heavy-tailed nature, is well-suited for modeling phenomena where rare but significant events have a large impact, such as the intense intermitent bursts observed in the waiting time distribution of repeating FRBs. Unlike the Weibull distribution, which is customized to model both clustered and spread-out events through its scale and shape parameters, the Pareto distribution effectively captures the sporadic nature of FRBs, where long periods of inactivity are interrupted by sudden, extreme bursts. By focusing on the occurrence of these irregular events, the Pareto distribution may provide a more fitting framework than non-Poisson distributions like the Weibull, particularly in accurately representing the extreme values denoted by heavy-tail is crucial to understanding FRBs dynamics.

Based on the previously established scale-invariant waiting time distribution Eq. \ref{scale_law}, $P$ (or $f$) \citep{duScalingUniversalityTemporal2024a}, this study investigates the temporal characteristics of the occurrences of FRB, specifically focusing on the distribution of time intervals between successive bursts in repeating FRBs. We analyze the conditional probability distribution $P_I(\lambda\vert \lambda_0)$, where $\lambda_0$ represents the interval between the last two bursts. For Poisson processes, which lack memory effects, $P_I(\lambda\vert \lambda_0)$ is expected to be identical to $P_I(\lambda)$ and independent of $\lambda_0$. However, our study aims to explore potential deviations from this behavior in repeating FRBs, which may indicate the presence of memory effects. We further quantify these temporal characteristics through various statistical measures, such as average time intervals and residual times under different conditions. These findings could provide insights into the probability of future burst occurrences based on existing observations of FRBs, contributing to a deeper understanding of the mechanisms governing FRB activity. Additionally, we propose a unified framework that might explain the waiting times of both repeating and apparently non-repeating FRBs.

\section{Method and Results} \label{sec:res}

For the purpose of studying the issue of temporal clustering of repeating FRBs, we search four high-quality datasets, The first dataset is FRB 20121102A, comprising $1652$ bursts detected over $59.5$ hours across $47$ days \citep{liBimodalBurstEnergy2021a}. The second dataset focuses on FRB 20201124A, featuring $1863$ bursts observed over $82$ hours spanning $54$ days \citep{xuFastRadioBurst2022a}. The third dataset is related to FRB 20220912A, which includes $1076$ bursts in $8.67$ hours over approximately $55$ days \citep{zhangFASTObservationsFRB2023}. The fourth dataset involves another extremely active episode of FRB 20201124A, containing $881$ bursts observed in approximately $4$ hours within a $4$-day period \citep{zhouFASTObservationsExtremely2022}. For the purpose of differentiation, in this paper, the two datasets related to FRB 20201124A in sequence of time will be denoted as FRB 20201124A(A) and FRB 20201124A(B) in the following sections. All datasets were recently acquired by FAST, which provided high raw sensitivity and temporal resolution detection \citep{qianFASTItsScientific2020}, effectively reducing the flux threshold by up to three times compared to previous observations for FRB 20121102A (refer to \cite{liBimodalBurstEnergy2021a} and the references cited therein). 
Recent research on FRB 20121102A and FRB 20201124A has revealed an absence of periodicity or quasiperiodicity on timescales ranging from $1$ ms to $1,000$ s \citep{liBimodalBurstEnergy2021a, xuFastRadioBurst2022a}.

\begin{figure}[!htb]
\centering
\includegraphics[width=0.9\textwidth]{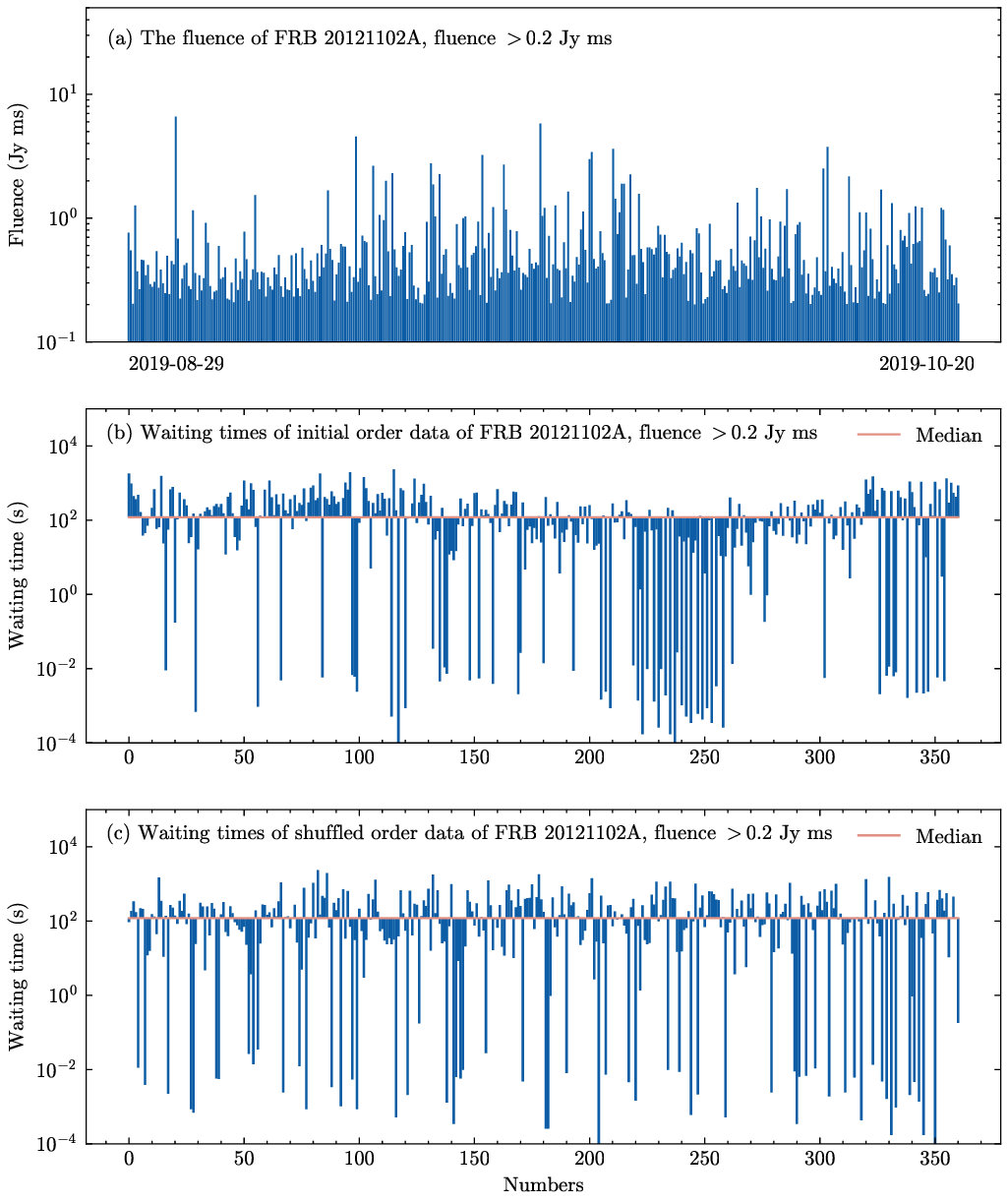}
\caption{(a) The FAST telescope observed the sequence of FRB 121102 with a fluence greater than 0.2 Jy ms during continuous observation sessions.(b) Sequence of the waiting time above and below the median $\lambda$ = 120 s (red line). (c) Same as (b) but for shuffled sequence of original data.}
\label{fig:seq}
\end{figure}

Due to the discontinuous nature of the observations conducted by FAST, we redefine the waiting time as the recurrence interval between consecutive events observed within the same observational period [see Fig. \ref{fig:seq}(a)]. In our analysis, we focus on waiting times, denoted
$\lambda$, for bursts with fluence exceeding a specified threshold value $I$. Here, $\lambda = t_{i+1} - t_i$, where $t_i$ and $t_{i+1}$ represent the arrival times of the $i$ th and $(i + 1)$ th bursts, respectively. Throughout our statistical calculations, we use logarithmic binning to ensure that each timescale is appropriately represented. Despite this, the calculation of the probability density function (PDF) makes the exact number and width of the bins less significant. In this context, the uncertainty associated with the counts of waiting times is estimated by $\sqrt{N_j}$, where $N_j$ represents the number of waiting times counted in the jth bin \citep{linScaleinvarianceRepeatingFast2019, weiSimilarScaleinvariantBehaviors2021a}.

As a typical example, the sequence of waiting times for FRB 20121102A is shown in Fig. \ref{fig:seq}(b), where the horizontal line is the median waiting time. The pattern of clusters of short and long waiting times, relative to the median, is apparent. In contrast, Fig. \ref{fig:seq}(c) displays the sequence of waiting times after shuffling the original sequence, and it notably lacks the similar clustering behavior observed in Fig. \ref{fig:seq}(b). The clusters in Fig. \ref{fig:seq}(b) qualitatively illustrate the presence of a memory effect in the waiting time sequence, where longer waiting times typically follow longer ones and shorter times follow shorter ones.

\begin{figure}[!htb]
\centering
\includegraphics[width=0.95\textwidth]{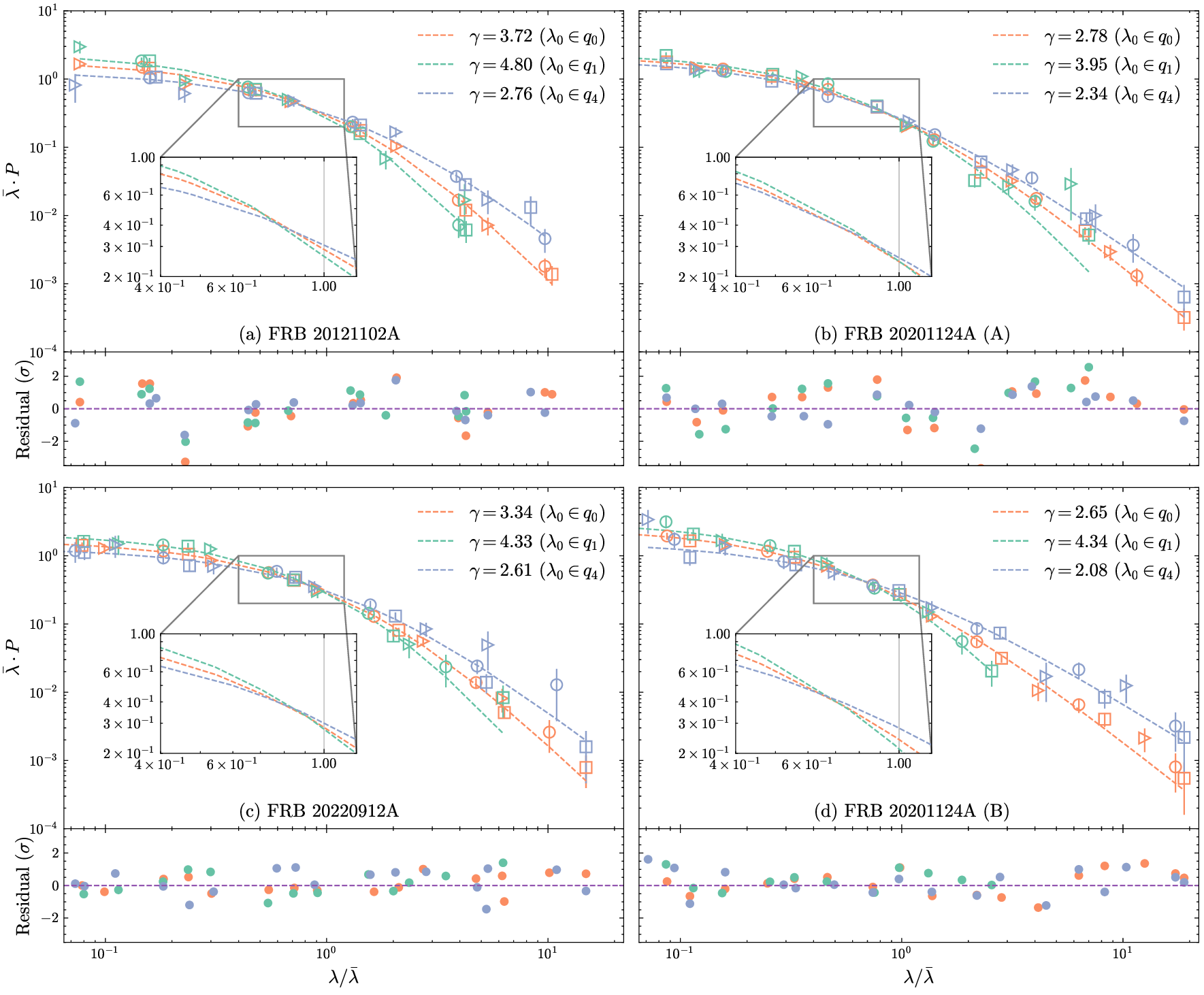}
\caption{The preconditioned distribution $P(\lambda \vert \lambda_0)$ characterizes the waiting times $\lambda$ between successive FRBs that exceed specified thresholds for various sources. The preceding condition values $\lambda_0$ are taken from both the first (minimal) and last (maximum) quartiles, denoted as $q_1$ and $q_4$, respectively. For the sake of uniform notation, $q_0$ is used to represent the entire dataset, implying that $P(\lambda \vert \lambda_0 \in q_0) = P(\lambda)$. (a) $P(\lambda \vert \lambda_0)$ for FRB 20121102A with fluence thresholds of 0.0 (square), 0.028 (circle) and 0.1 (right triangle) Jy ms; (b) $P(\lambda \vert \lambda_0)$ for FRB 20201124A(A) with fluence thresholds of 0.0 (square), 0.5 (circle), 1.0 (right triangle) Jy s; (c) $P(\lambda \vert \lambda_0)$ for FRB 20220912A with fluence thresholds of 0.0 (square), 0.1 (circle), 0.6 (right triangle) Jy ms; (d) $P(\lambda \vert \lambda_0)$ for FRB 20201124A(B) with fluence thresholds of 0.0 (square), 0.1 (circle), 0.5 (right triangle) Jy ms. For all subplots, the colors red, green, and blue correspond to $\lambda_0$ belonging to $q_0$, $q_1$, and $q_4$, respectively. The insets illustrate the behavior of the fitting curves near the point where $\lambda / \bar{\lambda} = 1$. The residuals between the data points and the best-fit values relative to the errors for all data are also displayed, with the color scheme corresponding to the different preconditions consistent with those in the preconditioned distribution $P(\lambda \vert \lambda_0)$.}
\label{fig:dist}
\end{figure}

To quantitatively characterize the memory effect, we examine the distribution of subsequent waiting times under different preceding conditions, that is, different values of the previous waiting time denoted as $\lambda_0$, represented as $P_I(\lambda \vert \lambda_0)$. To distinguish the magnitude of $\lambda_0$, the entire waiting time sequence is sorted in ascending order in time and divided into four subsets $q_1, q_2, q_3$ and $q_4$, each containing an equal number of elements. Fig. \ref{fig:dist} provides a comparative analysis of the waiting time distribution $\bar{\lambda}P_I(\lambda \vert \lambda_0)$ when the preceding condition $\lambda_0$ is placed in $q_1$ (green symbols) and $q_4$ (blue symbols). Furthermore, the distribution of waiting times without considering the preceding condition, $\bar{\lambda}P(\lambda)$, is plotted with the complete data sequence designated as $q_0$ (red symbols). In addition, different fluence thresholds are shown for different sources of FRB with various symbols, and the actual results are statistically independent of the magnitude of the fluence threshold. For the same FRB source and under the same preconditions, it can be observed that the waiting times have nearly the same distribution form under different fluence thresholds. The best-fit lines (the dashed lines) are obtained using the Markov Chain Monte Carlo (MCMC) method, and more detailed fitting information is listed in Table \ref{table:gamma}. From the table, we observe differences in the $\gamma$ values between certain datasets, especially for datasets with preconditions from $q_1$ and $q_4$, which exhibit a statistical significance approximately from $1.0$ to $3.5 \sigma$ (3.5, 0.94, 2.0, 3.2 for FRB 20121102A, FRB 20220912A, FRB 20201124A(A) and FRB 20201124A(B), respectively). This indicates a moderate to strong likelihood that the observed differences are not due to random chance, suggesting potential underlying variations between these comparisons of $q_1$ and $q_4$. However, the results of $0.94 \sigma$ and $2.0 \sigma$ are less significant and indicate that, for these particular comparisons, the evidence for the differences is weaker and could more likely be due to statistical fluctuations for limited samples. These mixed significance levels highlight the complexity of interpreting results with the current dataset size and the variability inherent in FRBs. Although some findings point to meaningful differences, they fall short of the more rigorous thresholds typically required for definitive conclusions. This underscores the need for caution when interpreting the results and highlights the importance of acquiring more extensive data. Furthermore, we compared the significance of the parameter $\gamma$ under different preconditions $(q_0, q_1, q_2, q_3, q_4)$ for FRB 20201124A(A) and FRB 20201124A(B). The significance levels were found to be $(0.59, 0.39, 0.19, 0.41, 0.80)$, indicating similarity within $1 \sigma$. Given that FRB 20201124A(B) occurred during a highly active period spanning 4 days, this similarity in $\gamma$ suggests an intrinsic connection in the burst behavior of the repeating FRB source during different active periods. This finding further supports the idea that the parameter $\gamma$ may capture underlying mechanisms that govern FRB activity, even across varying levels of source activity.

It is worth noting that a secondary peak in waiting time, ranging from a few to tens of milliseconds, was observed in the four burst source datasets, possibly due to the specific criteria used to characterize separate outbursts \citep{liBimodalBurstEnergy2021a, jahnsFRB20121102ANovember2023, xuFastRadioBurst2022a, zhangFASTObservationsFRB2023, zhouFASTObservationsExtremely2022}. Unlike the first peak, the location of the secondary peak remains constant as the fluence increases \citep{liBimodalBurstEnergy2021a}, disrupting the universal scaling relation \citep{duScalingUniversalityTemporal2024a}. As the occurrence of the first peak depends on the intensity of activity from the originating source and to ensure a uniform treatment across all distributions, we only consider the waiting times within the scaling region; that is, the minimal waiting time interval surpasses the secondary peak range, approximately $10$ s for FRB 20121102A and FRB 20201124A(A), and $1$ s for FRB 20220912A and FRB 20201124A(B).

As illustrated in Fig. \ref{fig:dist}, $P_I(\lambda \vert \lambda_0)$ generally exhibits an evident dependence on the preceding condition values $\lambda_0$, and changes its functional form accordingly. Later, when performing the specific calculations for a two-sample Kolmogorov-Smirnov (K-S) test, we will quantitatively discuss the difference of the distributions $P_I(\lambda \vert \lambda_0)$ again. Similarly to $P_I(\lambda)$, $P_I(\lambda \vert \lambda_0)$ demonstrates independence from the magnitude of fluence $I$. Hence, in subsequent analyzes,
the magnitude of fluence $I$ will be omitted unless explicitly needed. As observed in Fig. \ref{fig:dist}, when $\lambda_0$ is taken from $q_1$ and $q_4$, respectively, shorter waiting times are more probable ($P(\lambda \vert \lambda_0)$) in the interval $\lambda / \bar{\lambda} < 1$ compared to longer waiting times. In contrast, for $\lambda / \bar{\lambda} > 1$, the situation is reversed. Across the entire range, the $P(\lambda \vert \lambda_0)$ without any preceding condition ($\lambda_0 \in q_0$) falls between these two cases. 

Further fitting analysis reveals that $P(\lambda \vert \lambda_0)$ can also be well approximated by $P(\lambda \vert \lambda_0) \sim f(\lambda/\bar{\lambda}, \lambda_0/\bar{\lambda})/\bar{\lambda}$. Deviations from an exponential function can be well described by the Tsallis q-exponential function \citep{duScalingUniversalityTemporal2024a}. The parameter $\gamma$ in Eq. (\ref{g_distribution}) plays a key role in demonstrating the memory effects that the preceding condition values $\lambda_0$ have on subsequent FRB bursts and their degree of clustering. More details about $\gamma$ that are affected by different preceding conditions are provided in Table \ref{table:gamma}. For preconditioned intervals from $q_1$, $q_2$, $q_3$ to $q_4$, the parameter $\gamma$ generally decreases monotonically, a pattern that holds across the four datasets. However, considering the presence of fitting errors, there is a possible likelihood that overlapping intervals for $\gamma$ could occur between adjacent preconditioned intervals, such as between $[q_1, q_2]$, $[q_2, q_3]$, and $[q_3, q_4]$, and we also cannot completely rule out the possibility of overlapping intervals between $[q_1, q_4]$. After all, current statistical data samples are limited, and grouping further reduces the number of samples, which may lead to fluctuations in the statistical results. Furthermore, if the uncertainty of waiting times counts in the time intervals is considered, the parameter
$\gamma$ may differ compared to scenarios where this uncertainty is not taken into account. However, a precise determination of the values of the model parameter is not the main objective of this study. Instead, we primarily explore how different preconditioning, such as $q_1$, $q_2$, $q_3$, or $q_4$, influences the time intervals between subsequent FRB bursts, essentially studying the memory effect.

\begin{table}
\caption{Best-fitting values of the parameter $\gamma$, its $1\sigma$ errors and reduced Chi-square of the waiting time distribution $f(\lambda \vert \lambda_0)$, preconditioned or not. $q_1$, $q_2$, $q_3$ and $q_4$ means $\lambda_0$ is from the first, second, third and the last part of sorted data, while $q_0$ is the unconditional case.}
\begin{tabular}{c c c c c c} \hline
 & FRB 20121102A & FRB 20220912A & FRB 20201124A(A) & FRB 20201124A(B)\\
 & $\gamma$ $\hspace{1.0 cm}$ $\chi^2$ & $\gamma$ $\hspace{1.0 cm}$ $\chi^2$ & $\gamma$ $\hspace{1.0 cm}$ $\chi^2$ & $\gamma$ $\hspace{1.0 cm}$ $\chi^2$ & Mean of $\gamma$\\
\hline
$q_1$ & $4.80^{+0.37}_{-0.52}$ $\hspace{0.1 cm}$ 1.38 & $4.33^{+2.38}_{-1.18}$ $\hspace{0.1 cm}$ 0.77 & $3.95^{+0.97}_{-0.57}$ $\hspace{0.1 cm}$ 2.31 & $4.34^{+0.58}_{-0.71}$ $\hspace{0.1 cm}$ 0.48 & $4.26$\\
$q_2$ & $4.38^{+0.23}_{-0.19}$ $\hspace{0.1 cm}$ 1.23 & $4.27^{+1.80}_{-0.98}$ $\hspace{0.1 cm}$ 0.60 & $3.06^{+0.57}_{-0.37}$ $\hspace{0.1 cm}$ 1.12 & $3.19^{+0.62}_{-0.41}$ $\hspace{0.1 cm}$ 0.49 & $3.73$\\
$q_3$ & $3.24^{+0.31}_{-0.25}$ $\hspace{0.1 cm}$ 1.05 & $2.66^{+0.46}_{-0.36}$ $\hspace{0.1 cm}$ 0.78 & $2.92^{+0.37}_{-0.27}$ $\hspace{0.1 cm}$ 0.93 & $2.76^{+0.17}_{-0.27}$  $\hspace{0.1 cm}$ 0.52 & $2.90$\\
$q_4$ & $2.76^{+0.51}_{-0.24}$ $\hspace{0.1 cm}$ 0.75 & $2.61^{+0.69}_{-0.15}$ $\hspace{0.1 cm}$ 0.65 & $2.34^{+0.20}_{-0.15}$ $\hspace{0.1 cm}$ 0.59 & $2.08^{+0.31}_{-0.23}$ $\hspace{0.1 cm}$ 0.80 & $2.45$\\
$q_0$ & $3.72^{+0.32}_{-0.23}$ $\hspace{0.1 cm}$ 2.17 & $3.34^{+0.29}_{-0.24}$ $\hspace{0.1 cm}$ 0.33 & $2.78^{+0.12}_{-0.11}$ $\hspace{0.1 cm}$ 2.39 & $2.65^{+0.21}_{-0.17}$ $\hspace{0.1 cm}$ 0.67 & $3.12$\\
\hline
\end{tabular}
\label{table:gamma}
\end{table}

\begin{figure}[!htb]
\centering
\includegraphics[width=0.9\textwidth]{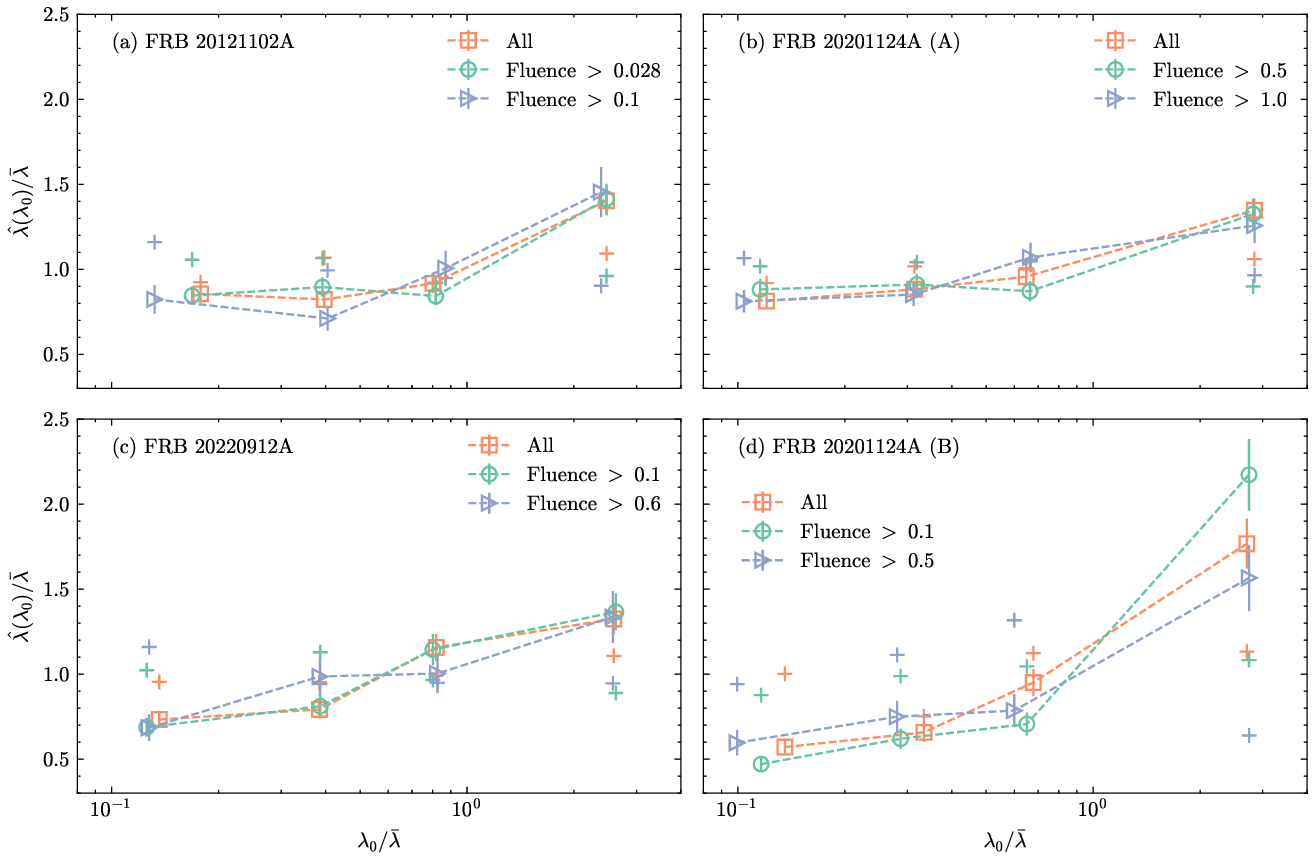}
\caption{The average value of conditional probability of waiting time $\hat{\lambda}(\lambda_0)/\bar{\lambda}$. They have similar form as $\lambda_0/\bar{\lambda}$ increasing with the same fluence thresholds shown in Fig. \ref{fig:dist}. The symbols of plus correspond to the shuffled sequences of waiting times which are fluctuating across the $\hat{\lambda}(\lambda_0)/\bar{\lambda} \approx 1$, indicative of a lack of memory in the data. }
\label{fig_tao}
\end{figure}

To quantify the impact of different preceding conditions $\lambda_0$ on subsequent burst events, we proceed to calculate the average value of the following burst waiting time distribution for every $\lambda_0 \in [q_1, q_2, q_3, q_4]$, denoted $\hat{\lambda}(\lambda_0)$. From Eq. (\ref{scale_law}), we obtain a fluence independent functional form $\hat{\lambda}(\lambda_0) = \int{\lambda P(\lambda \vert \lambda_0) d\lambda} \sim \bar{\lambda}g(\lambda_0/\bar{\lambda})$, where $g$ has a similar form for different FRBs. Fig. \ref{fig_tao} shows the ratio
$\hat{\lambda}(\lambda_0)/\bar{\lambda}$ as a function of $\lambda_0/\bar{\lambda}$. Due to the memory effect in the system, the value of $\hat{\lambda}/\bar{\lambda}$ is evidently below $1.0$ when $\lambda_0/\bar{\lambda}$ is close to $0.1$ and notably above $1.0$ when $\lambda_0/\bar{\lambda}$ is close to 100. This implies that shorter waiting times are more likely to be followed by shorter ones and longer waiting times by longer ones. For comparison, when the same preceding condition $\lambda_0$ is applied after random shuffling, the value of $\hat{\lambda}/\bar{\lambda}$ is always approximately $1.0$ (plus symbols in Fig. \ref{fig_tao}), indicating that the memory effect has been disrupted. Specifically, we note that FRB 20201124A(A) shows no significant difference from shuffled sequences, suggesting a lack of strong memory effects. In contrast, FRB 20201124A(B) probably exhibits more pronounced memory effects due to the intense burst activity over a short period, where such effects may be more evident. This variability highlights the complexity of FRB behavior, which may depend on the activity period and other unknown factors.

\begin{figure}[!htb]
\centering
\includegraphics[width=0.9\textwidth]{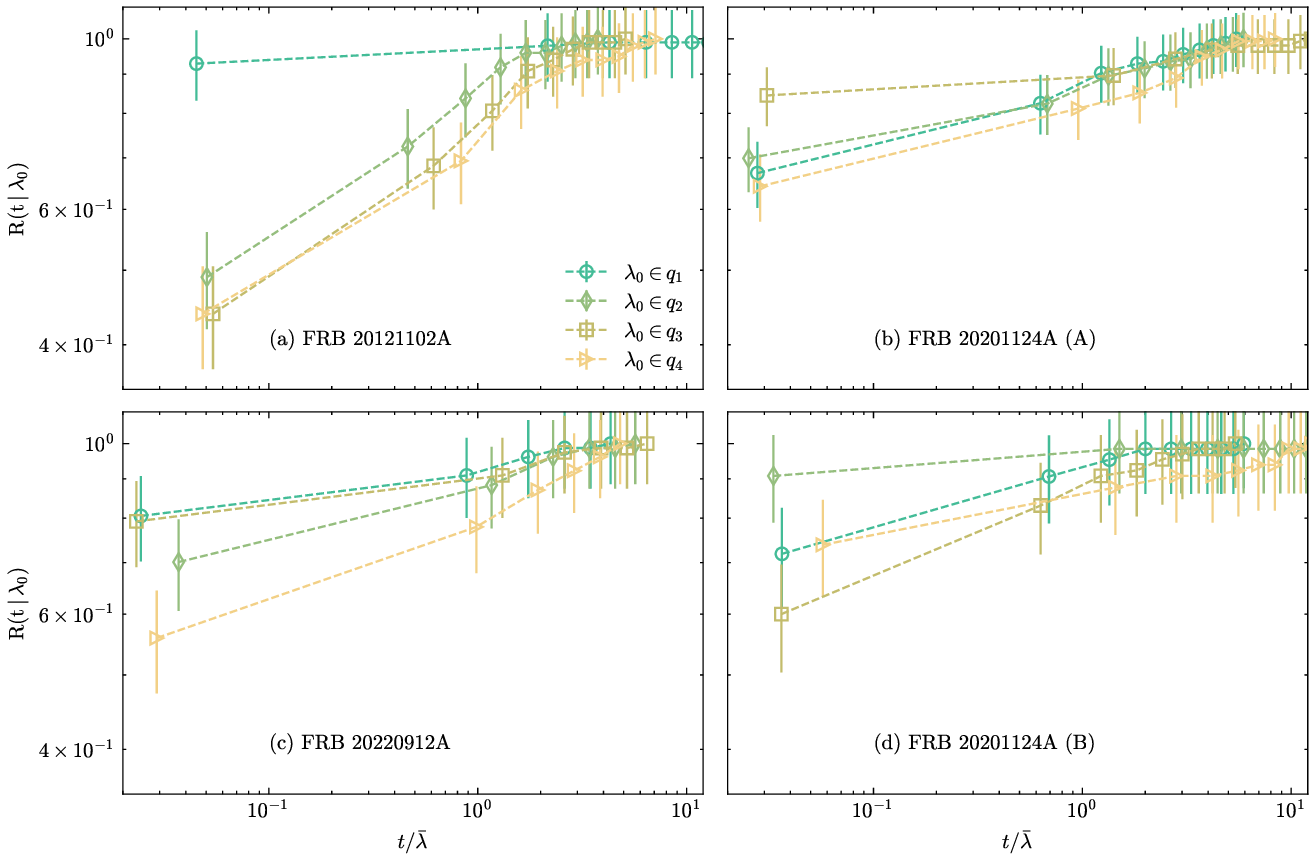}
\caption{The conditional probability of waiting time $R(t \vert \lambda_0)$, exceeding a specified fluence threshold for four FRBs datasets respectively, and separated by previous time interval $\lambda_0$, are succeeded by the next FRB within time t. The $\lambda_0$ comes from either $q_1$, $q_2$, $q_3$ or $q_4$ and thresholds are 0.1, 0.6, 0.5 Jy ms and 1.0 Jy s for FRB 20121102A, FRB 20220912A, FRB 20201124A(B) and FRB 20201124A(A).}
\label{fig_4pt}
\end{figure}

The obtained $\hat{\lambda}(\lambda_0)$ reflects the average time to wait for the next FRB burst under different preceding condition values $\lambda_0$. If we want to know more detailed temporal information about the next burst when the time interval between the previous two bursts is $\lambda_0$, this can be obtained by calculating the probability of the next burst occurring within a certain time $t$ or the cumulative distribution of $P(\lambda)$ (Eq. \ref{scale_law}), denoted $R(t \vert \lambda_0)$. Here, $t$ is a specific moment and the integral is conducted in the range of $(0,  t)$. This normalized probability reflects the distribution probabilities of the subsequent burst intervals in a different time range ($< t/\bar{\lambda}$) when the precondition is [$q_1, q_2, q_3, q_4$]. As $t$ becomes sufficiently large, the probability tends towards 1. Fig. \ref{fig_4pt} compares the probability distribution of the timing of the next burst when the time interval $\lambda_0$ is between the previous two bursts of $[q_1, q_2, q_3, q_4]$. For FRB 20121102A (Fig. \ref{fig_4pt}(a)) and FRB 20220912A (Fig. \ref{fig_4pt}(c)), the value of $\lambda_0$ clearly influences the timing of the next burst. This impact is particularly evident in the results $q_1$ and $q_4$, where the differences are especially pronounced. For FRB 20201124A(A) (Fig. \ref{fig_4pt}(b)) and FRB 20201124A(B) (Fig. \ref{fig_4pt}(d)) during different active periods, $R(t \vert \lambda_0)$ is relatively less affected by the magnitude of the values of the preceding conditions $\lambda_0$.

We conducted a two-sample K-S test comparing the differential distributions $P(\lambda|\lambda_0)$ where the preceding condition values $\lambda_0$ come from $q_1$ and $q_4$. The K-S statistics and P-values for FRB 20121102A, FRB 20220912A, FRB 20201124A(A) and FRB 20201124A(B) are as follows: (0.38, 0.14), (0.39, 0.11),  (0.22, 0.78) and (0.37, 0.19). All p-values are greater than the 0.05 threshold, indicating that there are no statistically significant differences. However, the small size and discrete nature of the binned data likely contribute to the increase in P-values, potentially obscuring subtle differences between distributions that the K-S test might miss. Therefore, a larger sample is potentially needed to further investigate and perhaps arrive at a significant p-value. For the results of FRB 20201124A(A), it could be associated with the instability of these distributions when dealing with limited statistics, with the expectation of achieving improved statistical results as more observational data become available. Naturally, this quantity $R(t \vert \lambda_0)$ represents only the average effect of the distribution function in specific contexts and, to some extent, may also highlight the complexity of the FRB waiting time distribution, which may be crucial to understanding the bursting behavior of FRBs.

From the preceding analysis, it is statistically evident that the time intervals between the first two bursts in repeating FRBs generally exert an evident influence on the occurrence of subsequent bursts. A realistic question arises: If a time $t$ has elapsed since the last burst without a subsequent occurrence, what is the average waiting time until the next burst? To answer this, one must calculate the expected residual time $\hat{\lambda}(t \vert \lambda_0)$. For the special case $t = 0$, which means that there is no time elapsed since the last burst, $\hat{\lambda}(t = 0 \vert \lambda_0) = \hat{\lambda}(\lambda_0)$. Based on the definition of residual time and the known waiting time distribution (Eq. \ref{g_distribution}), one can deduce that
\begin{equation}
    \hat{\lambda}(t \vert \lambda_0) = \frac{\int_{t}^{\infty}(\lambda - t)P(\lambda \vert \lambda_0)\, d\lambda}{\int_t^\infty \,d\lambda P(\lambda \vert \lambda_0)\, d\lambda}. \label{eq_residual_form}
\end{equation}
For distribution $P(\lambda \vert \lambda_0)$ that satisfied the Eq. (\ref{g_distribution}), we can obtain
\begin{equation}
    \hat{\lambda}(t \vert \lambda_0) \sim \frac{(t + \beta)(\gamma - 1)}{\gamma^2 - 3\gamma + 2},
    \label{eq_residual}
\end{equation}
where the integral is conducted under the condition that $\gamma > 2.0$.

\begin{figure}[!htb]
\centering
\includegraphics[width=0.9\textwidth]{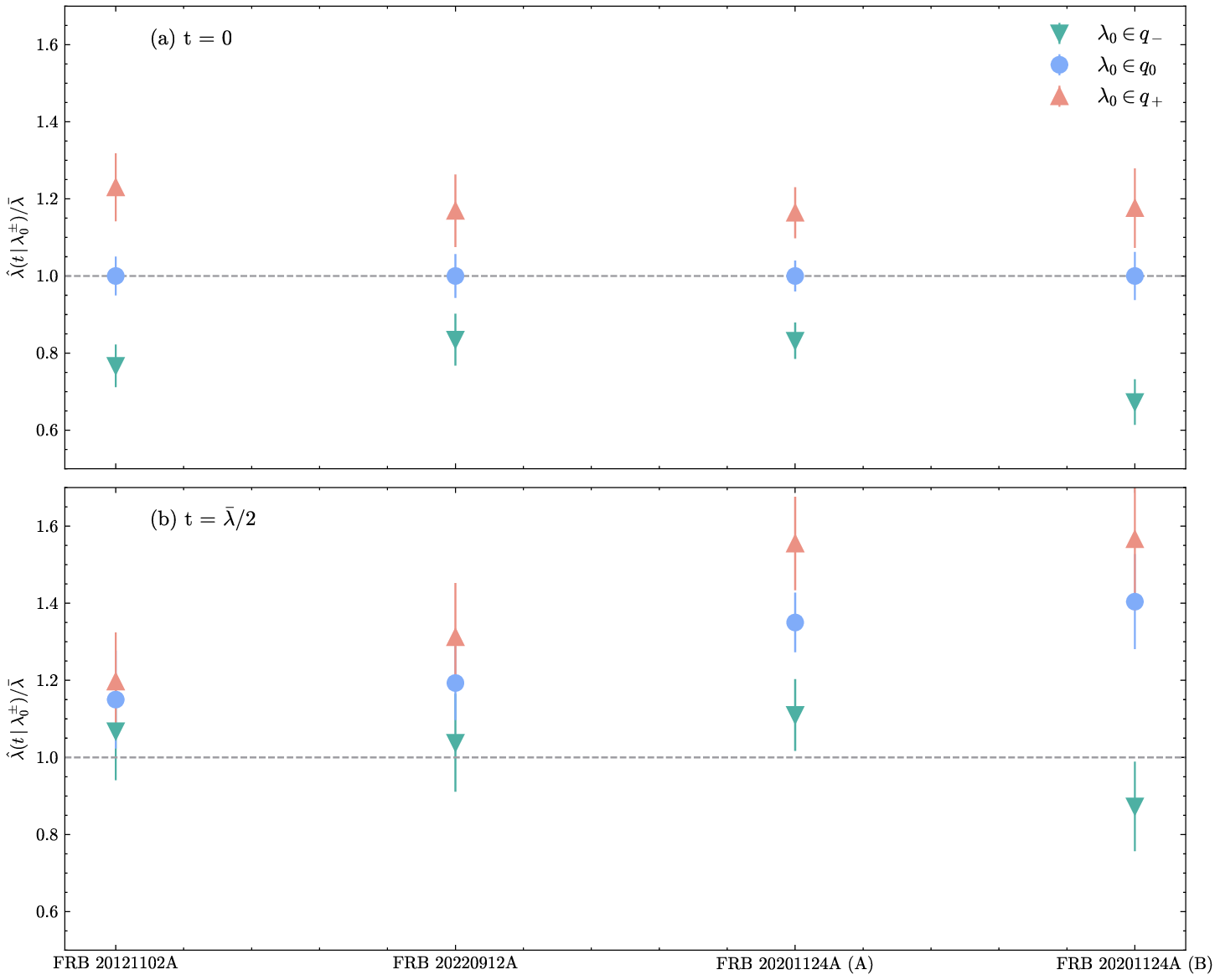}
\caption{The conditional residual time $\hat{\lambda}(t \vert \lambda_0)$. The $\lambda_0$ is either from $q_+$ (above median of entire data), $q_-$ (below median of entire data) or $q_0$ (the unconditional case). Two cases of $\hat{\lambda}(t \vert \lambda_0)$ for elapsed time t are considered for (a) $t = 0$ and (b) $t = \bar{\lambda}/2$. In the two cases all FRBs studied exhibit $\hat{\lambda}(t \vert \lambda_0^+) > \hat{\lambda}(t \vert \lambda_0^-)$, and are symmetrically distributed around $\hat{\lambda}(t)$ at $t = 0$. Additionally, relative to t = 0, the residual time $\hat{\lambda}(t \vert \lambda_0)$ at $t = \frac{\bar{\lambda}}{2}$ generally increases, regardless of whether $\lambda_0$ comes from $q_+$, $q_-$ or $q_0$. Fluences thresholds are the same as in Fig. \ref{fig_4pt}.}
\label{fig_ave}
\end{figure}

When there is no memory effect, $\hat{\lambda}(t \vert \lambda_0)$ is independent of both $\lambda_0$ and $t$. Conversely, if a memory effect exists, a clear dependency will be observed. The uncertainty in $\hat{\lambda}(t \vert \lambda_0)$ arises from the statistical error in the number of waiting time data \citep{linScaleinvarianceRepeatingFast2019, weiSimilarScaleinvariantBehaviors2021a} measured at $t = 0$ or other values. This uncertainty is then propagated into the calculation of the residual time. To quantitatively characterize the dependency of $\hat{\lambda}$ on $\lambda_0$ and $t$, the following scenarios were considered: First, the entire waiting time dataset is divided into $q_+$ and $q_-$, respectively, which is greater and smaller than its median, and $\lambda_0$ is taken from the intervals $q_+$ and $q_-$, denoted as $\lambda^+_0$ and $\lambda^-_0$. Furthermore, $\hat{\lambda}(t)$ without any preceding condition is also considered; Second, for comparison, $t$ is set to $0$ or $\bar{\lambda}/2$. The results are shown in Fig. \ref{fig_ave}, where the samples are from the datasets of FRB 20121102A, FRB 20220912A, FRB 20201124A(A) and FRB 20201124A(B), respectively. For the case of $t = 0$ (Fig. \ref{fig_ave}(a)), since $\hat{\lambda}(0) = (\hat{\lambda}(0\vert \lambda^-_0) + \hat{\lambda}(0\vert \lambda^+_0))/2$, the results show that $\hat{\lambda}(0\vert \lambda^-_0)$ (solid green triangle) and $\hat{\lambda}(0\vert \lambda^+_0)$ (solid red triangle) are approximately symmetrical to $\hat{\lambda}(0)$ (solid blue circle) $= 1$ (relative to $\bar{\lambda}$), and due to the memory effect, $\hat{\lambda}(0\vert \lambda^+_0)$ is significantly greater than $\hat{\lambda}(0\vert \lambda^-_0)$. The statistical significance of these differences in this case is $(4.5, 2.9, 4.1, 4.2)$ for FRB 20121102A, FRB 20220912A, FRB 20201124A(A) and FRB 20201124A(B), respectively.  Similar memory effects as shown in Fig. \ref{fig_ave}(a) are also evident in Fig. \ref{fig_ave}(b), that is, $\hat{\lambda}(\frac{\bar{\lambda}}{2}\vert \lambda^+_0)$ is generally greater than $\hat{\lambda}(\frac{\bar{\lambda}}{2}\vert \lambda^-_0)$. In this case, the significance of these differences is (0.8, 1.5, 2.9, 3.2). Compared to the situation at $t=0$, the reduction in significance may primarily be due to the increased statistical fluctuations caused by the reduced sample size. Furthermore, due to the memory effect, since the time elapsed is $t = \bar{\lambda}/2$ for the last burst was introduced, the symmetry between the residual times as shown in Fig. \ref{fig_ave}(a) is broken due to the arising of $\bar{\lambda}$, which is mostly contributed from $q_+$.

As illustrated in Fig. \ref{fig_ave}, after a time interval $t$ has elapsed since the last burst, the expected residual time to the next burst is generally greater than when $t = 0$. This can be directly inferred from Eq. \ref{eq_residual}. Generally, as $t$ increases, $\hat{\lambda}(t \vert \lambda_0)$ also increases monotonically, with the rate of increase influenced by the value of $\gamma$. Especially, for the case of $\gamma$ close to $2.0$ the $\hat{\lambda}(t \vert \lambda_0)$ is divergent. This behavior contrasts with the memoryless nature of the Poisson process \citep{sornetteParadoxExpectedTime1997, corralTimedecreasingHazardIncreasing2005, livinaMemoryOccurrenceEarthquakes2005}.

\section{Discussion}

Contrary to intuitive view, the probability that the next burst will occur does not increase with the time $t$ elapsed since the last burst; rather, it decreases, and the expected residual time to the next burst $\hat{\lambda}$ also increases. This serves as a more direct characterization of temporal clustering, where waiting times exhibit a tendency to cluster together.

An intuitive explanation for this phenomenon has shown that when a significant amount of time has elapsed since the last FRB burst, the system enters an enhanced ``period of inactivity'', during which the likelihood of an extended waiting time increases. However, it is important to note that these periods of inactivity are not fundamentally distinct from other waiting times, as they are all governed by the same continuous distribution. Furthermore, the fluence of the subsequent burst does not increase with the waiting time but appears to be independent of it.

Recent research, as referenced in \cite{aschwandenRECONCILIATIONWAITINGTIME2010}, offers a generalized understanding of non-stationary Poisson processes, which may be helpful in interpreting the results. Specifically, when the variability of the burst rate follows a gradual pattern and can be described by piecewise exponential functions, the exponent $\gamma$ in the waiting time distribution approximates 3.0. Conversely, if the burst rate variability exhibits sharp, intermittent spikes akin to $\delta$ functions, the exponent $\gamma$ in the waiting time distribution tends toward 2.0. This behavior, commonly referred to as ``clusterization'', is also found in earthquake and solar flare statistics \citep{corral_long_2004, corralTimedecreasingHazardIncreasing2005, Baiesi2006, aschwandenRECONCILIATIONWAITINGTIME2010}, which serves as indirect evidence for non-trivial space-time correlations among events. Consequently, the value of the exponent $\gamma$ serves as a critical indicator to determine whether the burst rate undergoes gradual or intermittent fluctuations. As a result, the bursts themselves are not periodic; the activity may occur within periodic windows \citep{weiPeriodicOriginFast2022}. Some recent studies have sought to distinguish between periodic windowed behavior and mere event clusters through time cluster analysis \citep{denissenyaDistinguishingTimeClustering2021}. More in-depth studies have shown that a similar memory effect is also observed in earthquakes \citep{livinaMemoryOccurrenceEarthquakes2005}. If repeating FRBs exhibit burst characteristics that are similar, or partially similar, to those observed in earthquakes or solar flares, the phenomena of ``clusterization'' and ``memory effects'' observed in these events could be more easily comprehensible. Indeed, current research \citep{duScalingUniversalityTemporal2024a} indicates that the scaling properties of repeating FRBs over longer time intervals (specifically, those exceeding approximately 1 s) are similar to those observed in solar flares, which could indicate a shared nature in their fundamental processes. In contrast, FRBs characterized by shorter intervals, that is, less than about 1 second, exhibit characteristics that are more similar to those associated with the dynamics of earthquakes \citep{totaniFastRadioBursts2023}.

In this study, the analysis of four datasets from three FRB sources reveals the value of $\gamma$ in the waiting time distribution (Eq. (\ref{g_distribution})) ranges approximately between $2.65$ and $3.72$ (for $q_0$ in Table \ref{table:gamma}). For cases where $\gamma$ is close to $2.0$, we make an estimate based on Eq. (\ref{eq_residual}), when $\beta = 1.0, \gamma = 2.1$, the expected residual time to the next burst $\hat{\lambda}$ is roughly 10 times the magnitude of the elapsed time $t$ after scaling. Here, the actual value of each scale $t$ is related to the average waiting time under different fluence thresholds. If $\beta$ remains constant and $\gamma$ decreases to $2.01$, then $\hat{\lambda}$ will be roughly $100$ times $t$. As $\gamma$ gradually approaches $2.0$, $\hat{\lambda}$ will become larger. The three burst sources analyzed in this paper have very different burst environments and properties. However, the statistical analysis results of these three repeating FRBs with the largest sample sizes indicate that the scale-invariant waiting time distribution and the clustering and memory of bursts are universal. This may be well used to explain the burst event characteristics of repeating FRBs that have been active in the past, as well as to explain the burst event characteristics that have only been observed once but may be repeating bursts. Although the existence of true one-offs is not ruled out, Eq. (\ref{eq_residual}) has also shown that when $\gamma = 2.0$, the expected residual time diverges and tends to be infinite, and this corresponds to the case of spikes like $\delta$ function of time dependent burst rate \citep{aschwandenRECONCILIATIONWAITINGTIME2010}, which mimics a true one-off burst from an observational perspective. Hence, Eq. (\ref{g_distribution}) offers a comprehensive explanation of the statistical characteristics of the waiting time distribution, accommodating the behavior of both repeating and one-off FRBs.

The complexity of waiting time between FRBs suggests that the physical models proposed thus far may not fully capture the complexity of the phenomenon. Currently, more than 700 FRBs have been observed, of which more than 60 exhibit repeating bursts \citep{Blinkverse}. Some analyses suggest that one-off and repeating FRBs have some statistical differences \citep{10.3847/1538-4357/ac958a, 10.1093/mnras/staa3351}. There are also studies discussing whether repeating and non-repeating FRBs have the same burst mechanisms \citep{calebAreAllFast2019a, 10.1038/s41586-020-2828-1, zhangPhysicsFastRadio2023, kirstenLinkRepeatingNonrepeating2024}, and partially explaining the statistical differences between one-off and repeating FRBs from an observational perspective \citep{iwazakiSpectraltemporalFeaturesRepeating2021}. However, more research is needed to determine the precise mechanisms that lead to the intermittent nature of FRBs and the extended periods between bursts. The ubiquity of this behavior implies the presence of a straightforward mechanism. In this mechanism, as time progresses, the variable responsible for triggering the burst increasingly deviates from the burst threshold on average. These ``fluctuations'' in the variable effectively preserve the memory of the last event within the system for extended periods, thereby contributing to the observed absence of time dependent aging. It is important to emphasize that these findings are in fact model independent, which means that they do not rely on any preconceived model of burst occurrence. Bak et al. \citep{Bak1996, bakUnifiedScalingLaw2002} suggests that the observed scaling law in burst events reflects a complex spatiotemporal pattern. This may be interpreted as an intermittent energy release from a self-organized system , which is consistent with the theoretical constructs of critical phenomena. Analogous memory effects, attributable to long-term correlations, have also been recently identified in the statistical analysis of two FRBs \citep{wangRepeatingFastRadio2023}, which address the self-similarity inherent in these processes, the framework of SOC currently offers the most compelling approach.

Our findings highlight general trends, but the limited sample size of independent repeating FRBs means that some exceptions may arise, and our conclusions should be interpreted with caution. The rarity and variability of FRBs underscore the need for more extensive datasets to validate and refine these preliminary insights. Advancements in detection technologies will likely provide the larger datasets needed to support our initial findings and enable more detailed studies of the characteristics and effects of FRB memory. Future observations will be crucial for verifying these trends, refining our understanding of FRB behavior, and uncovering significant patterns that may not yet be fully apparent. Continued research and data collection are essential for advancing our knowledge and confirming or challenging these early findings, ultimately deepening our understanding of these cosmic phenomena.

\section{Summary}
In conclusion, we have analyzed the statistics of waiting times between bursts (above certain fluence threshold values $I$) of three repeating FRBs. Our observations reveal a notable memory effect in these events, wherein small waiting times tend to follow small ones, and large waiting times follow large ones. This clustering of FRBs has been quantified using four metrics: (i) conditional distribution function, (ii) conditional mean recurrence time, (iii) conditional risk function, and (iv) conditional residual time. Each of these metrics generally deviates from their respective unconditional counterparts. The memory effect proves particularly beneficial as it enables us, as demonstrated in Fig. \ref{fig_ave}, to utilize relevant elapsed time data since the last burst to achieve a more accurate burst estimation. For both repeating FRBs in their active periods and the apparent non-repeating FRBs that have been observed to burst only once, a unified description of waiting time is provided. As a natural conclusion from the above results: Even if all FRBs are essentially repeating FRBs, the longer it has been since we have seen from those currently classified as one-off FRBs, the longer we may have to wait to see another burst, even if it is a repeating FRB. Conversely, the more recently observed one-off FRBs are more likely to be seen bursting again in a shorter period of time.

\section*{Data availability}
All data analyzed in this article are included in references \citep{liBimodalBurstEnergy2021a, xuFastRadioBurst2022a, zhangFASTObservationsFRB2023, zhouFASTObservationsExtremely2022} separately. They are also openly available in the Science Data Bank at \dataset{https://doi.org/10.11922/sciencedb.01092} \doi{10.11922/sciencedb.01092} \citep{liBimodalBurstEnergy2021a}, \dataset{https://cstr.cn/31253.11.sciencedb.08058} \doi{10.57760/sciencedb.08058} \citep{zhangFASTObservationsFRB2023}, \dataset{https://cstr.cn/31253.11.sciencedb.06762} \doi{10.57760/sciencedb.06762} \citep{zhouFASTObservationsExtremely2022}, and \dataset{https://doi.org/10.6084/m9.figshare.19688854.v2} \doi{10.6084/m9.figshare.19688854.v2} \citep{xuFastRadioBurst2022a}.


\begin{acknowledgments}
\textbf{Acknowledgements:} We thank Mingyu Ge, Sai Wang, and Xiaobo Li for useful discussions. S.-L.X. acknowledge the support of the National Natural Science Foundation of China (Grant No. 12273042), S.X. acknowledge the support
by the National Natural Science Foundation of China (NO. 12303043) and C.C. acknowledge the support
by the Science Foundation of Hebei Normal University (No. L2023B11), the National Natural Science Foundation of China (No. 12303045) and the Natural Science Foundation of Hebei Province (No. A2023205020). The authors acknowledge the computing resources provided by the National High Energy Physics Data Center and the High Energy Physics Data Center of the Chinese Academy of Science.
\end{acknowledgments}




\bibliography{article}{}
\bibliographystyle{aasjournal}



\end{document}